\documentclass[preprint,onecolumn,aps,pre,eqsecnum,longbibliography,showpacs,showkeys,a4paper,nofootinbib]{revtex4-1}
\usepackage[latin9]{inputenc}
\usepackage{color}
\definecolor{note_fontcolor}{rgb}{0.800781, 0.800781, 0.800781}
\usepackage[british]{babel}
\usepackage{units}
\usepackage{amsmath}
\usepackage{amssymb}
\usepackage{graphicx}
\usepackage[unicode=true,
 bookmarks=false,
 breaklinks=true,pdfborder={0 0 0},pdfborderstyle={},backref=false,colorlinks=true]
 {hyperref}
\hypersetup{
 pdfauthor={AINS}}

\makeatletter

\newenvironment{lyxgreyedout}
  {\textcolor{note_fontcolor}\bgroup\ignorespaces}
  {\ignorespacesafterend\egroup}

\numberwithin{equation}{section}
\numberwithin{figure}{section}
\numberwithin{table}{section}

\usepackage{graphicx}

\DeclareGraphicsExtensions{.png,.pdf,.eps}
\graphicspath{{pictures/}}

\makeatother

\begin{document}


\title{Vacuum Landscaping: Cause of Nonlocal Influences without Signalling}

\author{Gerhard \surname{Grössing}\textsuperscript{}}
\email[Corresponding author: ]{ains@chello.at}

\homepage{http://www.nonlinearstudies.at}

\selectlanguage{british}%

\author{Siegfried \surname{Fussy}}

\author{Johannes \surname{Mesa Pascasio}}

\author{Herbert \surname{Schwabl}}

\affiliation{Austrian Institute for Nonlinear Studies, Akademiehof, Friedrichstr.~10,
1010 Vienna, Austria\vspace{1cm}
}
\begin{abstract}
In the quest for an understanding of nonlocality with respect to an
appropriate ontology, we propose a \textquotedblleft cosmological
solution\textquotedblleft . We assume that from the beginning of the
universe each point in space has been the location of a scalar field
representing a zero-point vacuum energy that nonlocally vibrates at
a vast range of different frequencies across the whole universe. A
quantum, then, is a nonequilibrium steady state in the form of a \textquotedblleft bouncer\textquotedblleft{}
coupled resonantly to one of those (particle type dependent) frequencies,
in remote analogy to the bouncing oil drops on an oscillating oil
bath as in Couder's experiments. A major difference to the latter
analogy is given by the nonlocal nature of the vacuum oscillations.

We show with the examples of double- and $n$-slit interference that
the assumed nonlocality of the distribution functions alone suffices
to derive the de\textasciitilde{}Broglie-{}-Bohm guiding equation
for $N$ particles with otherwise purely classical means. In our model,
no influences from configuration space are required, as everything
can be described in 3-space. Importantly, the setting up of an experimental
arrangement limits and shapes the forward and osmotic contributions
and is described as vacuum landscaping. %
\begin{lyxgreyedout}
\global\long\def\VEC#1{\mathbf{#1}}

\global\long\def\d{\,\mathrm{d}}

\global\long\def\e{{\rm e}}

\global\long\def\i{{\rm i}}

\global\long\def\meant#1{\left<#1\right>}

\global\long\def\meanx#1{\overline{#1}}

\global\long\def\mpbracket{\ensuremath{\genfrac{}{}{0pt}{1}{-}{\scriptstyle (\kern-1pt +\kern-1pt )}}}

\global\long\def\pmbracket{\ensuremath{\genfrac{}{}{0pt}{1}{+}{\scriptstyle (\kern-1pt -\kern-1pt )}}}

\global\long\def\p{\partial}
\end{lyxgreyedout}
\end{abstract}
\maketitle

\section{Introduction: Quantum Mechanics without Wavefunctions\label{sec:introduction}}

``Emergent Quantum Mechanics'' stands for the idea that quantum
mechanics is based on a more encompassing deeper level theory. This
counters the traditional belief, usually expressed in the context
of orthodox Copenhagen-type quantum mechanics, that quantum theory
is an ``ultimate'' theory whose main features will prevail for all
time and will be applicable to all questions of physics. Note, for
example, that even in more recent approaches to spacetime, the concept
of an ``emergent spacetime'' is introduced as a description even
of space and time emerging from basic quantum mechanical entities.
This, of course, need not be so, considering the fact that there is
``plenty of room at the bottom'', i.e.\ as Feynman implied, between
present-day resolutions and minimally possible times and distances,
which could in principle be way below resolutions reasonably argued
about in present times (i.e.\ on Planck scales).

One of the main attractive features of the de~Broglie\textendash Bohm
interpretation of the quantum mechanical formalism, and of Bohmian
mechanics as well, lies in the possibility to extend its domain into
space and/or time resolutions where modified behaviours different
from quantum mechanical ones may be expected. In other words, there
may be new physics involved that would require an explicitly more
encompassing theory than quantum mechanics, i.e.\ a deeper level
theory. Our group's approach, which we pursued throughout the last
10 years, is characterized by the search for such a theory under the
premise that even for nonrelativistic quantum mechanics, the Schrödinger
equation cannot be an appropriate starting point, since the wavefunction
is still lacking a firm theoretical basis and its meaning is generally
not agreed upon.

For a similar reason, also the de~Broglie\textendash Bohm theory
cannot be our starting point, as it is based on the Schrödinger equation
and the use of the wavefunction to begin with. Rather, we aim at an
explicit ansatz for a deeper level theory without wavefunctions, from
which the Schrödinger equation, or the de~Broglie\textendash Bohm
guiding equation, can be derived. We firmly believe that we have accomplished
this and we can now proceed to study consequences of the approach
beyond orthodox expectations.

Throughout recent years, apart from our own model, several approaches
to a quantum mechanics without wavefunctions have been proposed~\cite{Deckert.2007quantum,Poirier.2010bohmian,Poirier.2012trajectory-based,Schiff.2012communication:,Hall.2014quantum}.
These refer to ``many classical worlds'' which provide Bohm-type
trajectories with certain repulsion effects. From our realistic point
of view, the true ontologies of these models, however, do not become
apparent. So let us turn to our model. As every physical theory is
based on metaphysical assumptions, we must make clear what our assumptions
are. Here they are.

We propose a ``cosmological solution'' in that the Big Bang, or
any other model explaining the apparent expansion of the universe,
is essentially related to the vacuum energy. (The latter may constitute
what is called the dark energy, but we do not need to specify this
here.) We assume that from the beginning of the universe each point
in space has been the location of a scalar field representing a zero-point
vacuum energy that vibrates at a vast range of different frequencies
across the whole universe. More specifically, we consider the universe
as an energetically open system where the vacuum energy not only drives
expansion, but also each individual ``particle'' oscillation $\omega=E/\hbar$
in the universe. In order to maintain a particular frequency, any
such oscillator must be characterized by a throughput of energy external
to it. In this regard, we have time and again employed the analogy
of Couder's experiments with bouncing oil drops on a vibrating bath~\cite{Couder.2005,Couder.2006single-particle,Couder.2012probabilities,Bush.2010quantum,Bush.2015new,Bush.2015pilot-wave}:
The bouncer/particle is always in resonant interaction with a relevant
environment.

Our model, though also largely classical, has a very different ontology
from the ``many classical worlds'' one. We consider \emph{one} ``superclassical''
world instead: a purely classical world plus ``cosmological nonlocality'',
i.e.\ a nonlocal bath for every oscillator/particle due to the all-pervading
vacuum energy, which \textendash{} mostly in the context of quantum
mechanics \textendash{} is called the zero-point energy. So, it is
the one classical world together with the fluctuating environment
related to the vacuum energy that enters our definition of a quantum
as an emergent system. The latter consists of a bouncer and an undulatory/wave-like
nonlocal environment defined by proper boundary conditions.\footnote{As an aside we note that this is not related to de~Broglie's ``nonlinear
wave mechanics''~\cite{DeBroglie.1960book}, as there the nonlinear
wave, with the particle as soliton-like singularity, is considered
as one ontic entity. In our case, however, we speak of two separate,
though synchronous elements: local oscillators and generally nonlocal
oscillating fields.}

In previous work, we have shown how the Schrödinger equation can be
derived from a nonequilibrium sub-quantum dynamics~\cite{Groessing.2008vacuum,Groessing.2010emergence,Groessing.2011dice,Groessing.2012doubleslit},
where in accordance with the model sketched above the particle is
considered as a steady state with a constant throughput of energy.
This, then, leads to the two-momenta approach to emergent quantum
mechanics which shall be outlined in the next section.

\section{The Two-Momenta Approach to Emergent Quantum Mechanics\label{sec:motivation}}

We consider the empirical fact that each particle of nature is attributed
an energy $E=\hbar\omega$ as one of the essential features of quantum
systems.\footnote{We have also presented a classical explanation for this relation from
our sub-quantum model~\cite{Groessing.2011explan}, but do not need
to use the details for our present purposes.} Oscillations, characterized by some typical angular frequency $\omega$,
are described as properties of off-equilibrium steady-state systems.
''Particles'' can then be assumed to be dissipative systems maintained
in a nonequilibrium steady-state by a permanent throughput of energy,
or heat flow, respectively. The heat flow must be described by an
external kinetic energy term. Then the energy of the total system,
i.e.\ of the particle and it's thermal context, becomes
\begin{equation}
E_{{\rm tot}}=\hbar\omega+\frac{(\delta\VEC p)^{2}}{2m}\;,\label{eq:2.1}
\end{equation}
where $\delta\VEC p$ is an additional, fluctuating momentum component
of the particle of mass $m$.

We assume that an effect of said thermal context is given by detection
probability distributions which are wave-like in the particle's surroundings.
Thus, the detection probability density $P(\VEC x,t)$ is considered
to coincide with a classical wave's intensity $I(\VEC x,t)=R^{2}(\VEC x,t),$
with $R(\VEC x,t)$ being the wave's real-valued amplitude
\begin{equation}
P(\VEC x,t)=R^{2}(\VEC x,t)\;,\quad\text{with normalization}\,\int P\d^{n}x=1\;.\label{eq:2.2}
\end{equation}

In ref.~\cite{Groessing.2008vacuum}, we combine some results of
nonequilibrium thermodynamics with classical wave mechanics. We propose
that the many microscopic degrees of freedom associated with the hypothesized
sub-quantum medium can be recast into the emergent macroscopic properties
of the wave-like behaviour on the quantum level. Thus, for the relevant
description of the total system one no longer needs the full phase
space information of all microscopic entities, but only the emergent
particle coordinates.

For implementation, we model a particle as being surrounded by a heat
bath, i.e.\ a reservoir that is very large compared to the small
dissipative system, such that that the momentum distribution in this
region is given by the usual Maxwell\textendash Boltzmann distribution.
This corresponds to a ``thermostatic'' regulation of the reservoir's
temperature, which is equivalent to the statement that the energy
lost to the thermostat can be regarded as heat. Thus, one can formulate
a \textit{proposition of emergence} \cite{Groessing.2008vacuum} providing
the equilibrium-type probability (density) ratio 
\begin{equation}
\frac{P(\VEC x,t)}{P(\VEC x,0)}=\e^{-\frac{\Delta Q(t)}{kT}}\;,\label{eq:2.3}
\end{equation}
with $k$ being Boltzmann's constant, $T$ the reservoir temperature,
and $\Delta Q(t)$ the heat that is exchanged between the particle
and its environment.

Equations (\ref{eq:2.1}), (\ref{eq:2.2}), and (\ref{eq:2.3}) are
the only assumptions necessary to derive the Schrödinger equation
from (modern) classical mechanics. We need to employ only two additional
well-known results. The first is given by Boltzmann's formula for
the slow transformation of a periodic motion (with period $\tau=2\pi/\omega$)
upon application of a heat transfer $\Delta Q$. This is needed as
we deal with an oscillator of angular frequency $\omega$ in a heat
bath $Q$, and a change in the vacuum surroundings of the oscillator
will come as a heat transfer $\Delta Q$ . The latter is responsible
for a change $\delta S$ of the action function $S$ representing
the effect of the vacuum's ``zero-point'' fluctuations. With the
action function $S=\int\left(E_{{\rm kin}}-V\right)\d t$, the relation
between heat and action was first given by Boltzmann~\cite{Boltzmann.1866uber},
\begin{equation}
\Delta Q(t)=2\omega[\delta S(t)-\delta S(0)]\;.\label{eq:2.4}
\end{equation}
Finally, the requirement that the average kinetic energy of the thermostat
equals the average kinetic energy of the oscillator is given, for
each degree of freedom, by
\begin{equation}
\frac{kT}{2}=\frac{\hbar\omega}{2}\;.\label{eq:2.5}
\end{equation}
Combining these two results, Eqs.~(\ref{eq:2.4}) and (\ref{eq:2.5}),
with~(\ref{eq:2.3}) one obtains 
\begin{equation}
P(\VEC x,t)=P(\VEC x,0)\e^{-\frac{2}{\hbar}[\delta S(\VEC x,t)-\delta S(\VEC x,0)]}\;,\label{eq:2.6}
\end{equation}
from which follows the expression for the momentum fluctuation $\delta\VEC p$
of (\ref{eq:2.1}) as 
\begin{equation}
\delta\VEC p(\VEC x,t)=\nabla(\delta S(\VEC x,t))=-\frac{\hbar}{2}\frac{\nabla P(\VEC x,t)}{P(\VEC x,t)}\;.\label{eq:2.7}
\end{equation}
This, then, provides the additional kinetic energy term for one particle
as
\begin{equation}
\delta E_{{\rm kin}}=\frac{1}{2m}\nabla(\delta S)\cdot\nabla(\delta S)=\frac{1}{2m}\left(\frac{\hbar}{2}\frac{\nabla P}{P}\right)^{2}\;.\label{eq:2.8}
\end{equation}
Thus, writing down a classical action integral for $j=N$ particles
in $m$-dimensional space, including this new term for each of them,
yields (with external potential $V$) 
\begin{equation}
A=\int L\d^{m}x\d t=\int P\left[\frac{\partial S}{\partial t}+\sum_{j=1}^{N}\frac{1}{2m_{j}}\nabla_{j}S\cdot\nabla_{j}S+\sum_{j=1}^{N}\frac{1}{2m_{j}}\left(\frac{\hbar}{2}\frac{\nabla_{j}P}{P}\right)^{2}+V\right]\d^{m}x\d t\;,\label{eq:2.9}
\end{equation}
where the probability density $P=P(\VEC x_{1},\VEC x_{2},\ldots,\VEC x_{N},t)$. 

With the definition of forward and osmotic velocities, respectively,
\begin{equation}
\VEC v_{j}:=\frac{\VEC p_{j}}{m_{j}}=\frac{\nabla_{j}S}{m_{j}}\qquad\textrm{ and }\qquad\VEC u_{j}:=\frac{\delta\VEC p_{j}}{m_{j}}=-\frac{\hbar}{2m_{j}}\frac{\nabla_{j}P}{P},\label{eq:2.9b}
\end{equation}
 one can rewrite (\ref{eq:2.9}) as
\begin{equation}
A=\int L\d^{m}x\d t=\int P\left[\frac{\partial S}{\partial t}+V+\sum_{j=1}^{N}\frac{m_{j}}{2}\VEC v_{j}^{2}+\sum_{j=1}^{N}\frac{m_{j}}{2}\VEC u_{j}^{2}\right]\d^{m}x\d t\;.\label{eq:2.9c}
\end{equation}
This can be considered as the basis for our approach with two momenta,
i.e.\ the forward momentum $m\VEC v$ and the osmotic momentum $m\VEC u$,
respectively. At first glance, the Lagrangian in Eq.~(\ref{eq:2.9c})
looks completely classical, with two kinetic energy terms per particle
instead of one. However, due to the particular nature of the osmotic
momentum as given in Eq.~(\ref{eq:2.9b}), nonlocal influences are
introduced: even at long distances away from the particle location,
where the particle's contribution to $P$ is practically negligibly
small, the expression of the form $\frac{\nabla_{j}P}{P}$ may be
large and affects immediately the whole fluctuating environment. This
is why the osmotic variant of the kinetic energy makes all the difference
to the usual classical mechanics, or, in other words, is the basis
for quantum mechanics.

Introducing now the Madelung transformation
\begin{equation}
\psi=R\;\e^{\frac{\i}{\hbar}S}\;,\label{eq:2.10}
\end{equation}
where $R=\sqrt{P}$ as in (\ref{eq:2.2}), one has, with bars denoting
averages, 
\begin{equation}
\overline{\left|\frac{\nabla_{j}\psi}{\psi}\right|^{2}}:=\int\d^{m}x\d t\left|\frac{\nabla_{j}\psi}{\psi}\right|^{2}=\overline{\left(\frac{1}{2}\frac{\nabla_{j}P}{P}\right)^{2}}+\overline{\left(\frac{\nabla_{j}S}{\hbar}\right)^{2}}\;,\label{eq:2.11}
\end{equation}
and one can rewrite (\ref{eq:2.9}) as 
\begin{equation}
A=\int L\d^{m}x\d t=\int\d^{m}x\d t\left[|\psi|^{2}\left(\frac{\partial S}{\partial t}+V\right)+\sum_{j=1}^{N}\frac{\hbar^{2}}{2m_{j}}|\nabla_{j}\psi|^{2}\right]\;.\label{eq:2.12}
\end{equation}
Thus, with the identity $|\psi|^{2}\frac{\partial S}{\partial t}=-\frac{\i\hbar}{2}(\psi^{*}\dot{\psi}-\dot{\psi}^{*}\psi)$,
one obtains the familiar Lagrange density 
\begin{equation}
L=-\frac{\i\hbar}{2}(\psi^{*}\dot{\psi}-\dot{\psi}^{*}\psi)+\sum_{j=1}^{N}\frac{\hbar^{2}}{2m_{j}}\nabla_{j}\psi\cdot\nabla_{j}\psi^{*}+V\psi^{*}\psi\;,\label{eq:2.13}
\end{equation}
from which by the usual procedures one arrives at the $N$-particle
Schrödinger equation 
\begin{equation}
\i\hbar\frac{\partial\psi}{\partial t}=\left(-\sum_{j=1}^{N}\frac{\hbar^{2}}{2m_{j}}\nabla_{j}^{2}+V\right)\psi\;.\label{eq:2.14}
\end{equation}
Note also that from (\ref{eq:2.9}) one obtains upon variation in
$P$ the modified Hamilton\textendash Jacobi equation familiar from
the de~Broglie\textendash Bohm interpretation, i.e.
\begin{equation}
\frac{\partial S}{\partial t}+\sum_{j=1}^{N}\frac{(\nabla_{j}S)^{2}}{2m_{j}}+V(\VEC x_{1},\VEC x_{2},\dots,\VEC x_{N},t)+U(\VEC x_{1},\VEC x_{2},\ldots,\VEC x_{N},t)=0\;,\label{eq:2.15}
\end{equation}
where $U$ is known as the ``quantum potential'' 
\begin{equation}
U(\VEC x_{1},\VEC x_{2},\ldots,\VEC x_{N},t)=\sum_{j=1}^{N}\frac{\hbar^{2}}{4m_{j}}\left[\frac{1}{2}\left(\frac{\nabla_{j}P}{P}\right)^{2}-\frac{\nabla_{j}^{2}P}{P}\right]=-\sum_{j=1}^{N}\frac{\hbar^{2}}{2m_{j}}\frac{\nabla_{j}^{2}R}{R}\;.\label{eq:2.16}
\end{equation}
Moreover, with the definitions of $\VEC u_{j}$ in~(\ref{eq:2.9b})
one can rewrite $U$ as 
\begin{equation}
U=\sum_{j=1}^{N}\left[\frac{m_{j}\VEC u_{j}^{2}}{2}-\frac{\hbar}{2}(\nabla_{j}\cdot\VEC u_{j})\right]\;.\label{eq:2.18}
\end{equation}
However, as was already pointed out in ref.~\cite{Groessing.2008vacuum},
with the aid of (\ref{eq:2.4}) and (\ref{eq:2.6}), $\VEC u_{j}$
can also be written as 
\begin{equation}
\VEC u_{j}=\frac{1}{2\omega_{j}m_{j}}\nabla_{j}Q\;,\label{eq:2.19}
\end{equation}
which thus explicitly shows its dependence on the spatial behaviour
of the heat flow $\delta Q$. Insertion of (\ref{eq:2.19}) into (\ref{eq:2.18})
then provides the thermodynamic formulation of the quantum potential
as 
\begin{equation}
U=\sum_{j=1}^{N}\frac{\hbar^{2}}{4m_{j}}\left[\frac{1}{2}\left(\frac{\nabla_{j}Q}{\hbar\omega_{j}}\right)^{2}-\frac{\nabla_{j}^{2}Q}{\hbar\omega_{j}}\right]\;.\label{eq:2.20}
\end{equation}

\begin{figure}[h]
\centering{}%
\begin{minipage}[t]{0.9\columnwidth}%
\begin{center}
\includegraphics[width=1\columnwidth]{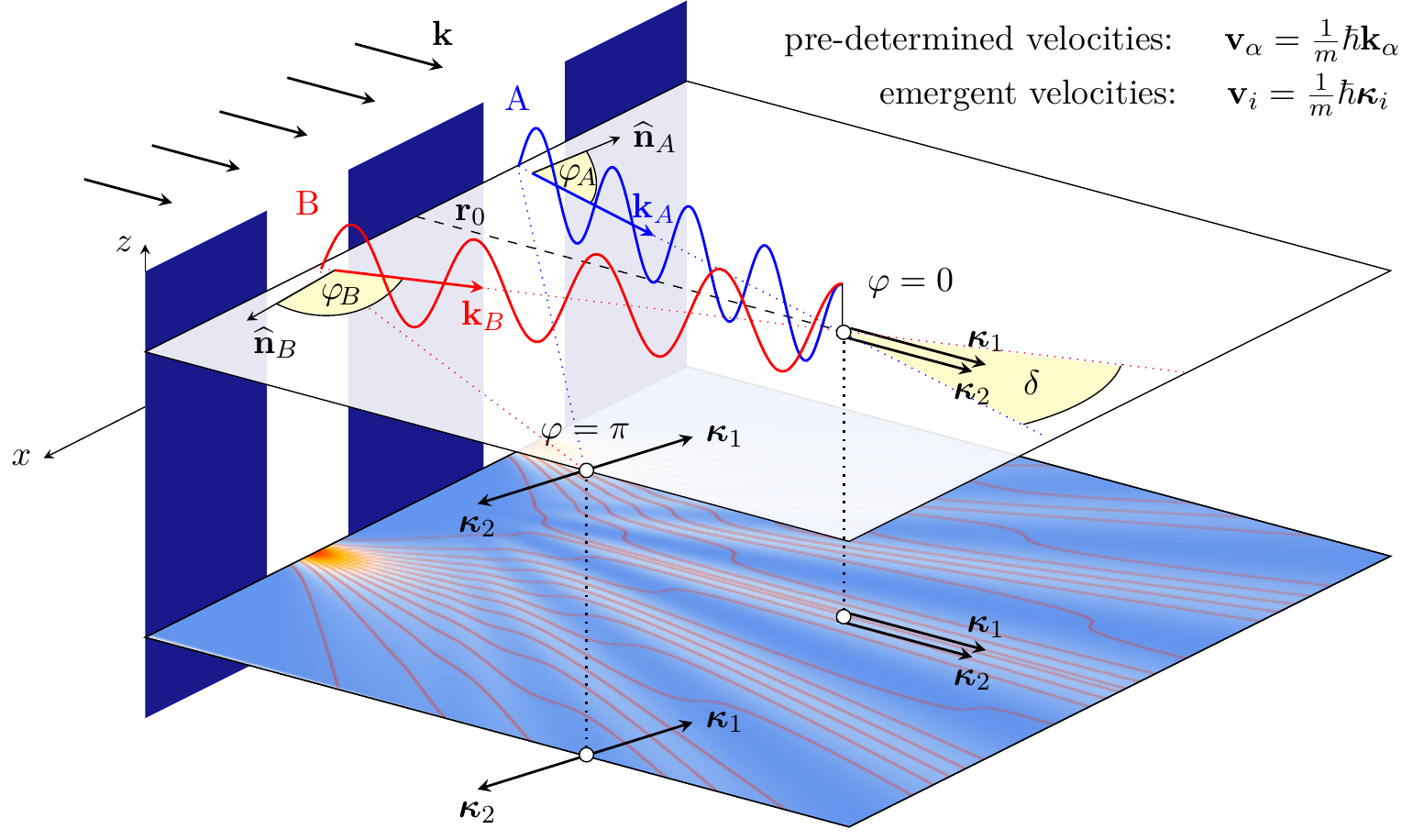}\caption{{\small{}Scheme of interference at a double-slit. Considering an incoming
beam of electrons with wave number $\mathbf{k}$ impinging on a wall
with two slits, two beams with wave numbers $\mathbf{k}_{A}$ and
$\mathbf{k}_{B}$, respectively, are created, which one may denote
as ``pre-determined'' velocities $\mathbf{v}_{\alpha}=\frac{1}{m}\hbar\mathbf{k}_{\alpha}\textrm{,\:\ensuremath{\alpha}=\ensuremath{A}\:or\:\ensuremath{B}.}$
Taking into account the influences of the osmotic momentum field $m\mathbf{u}$,
one has to combine all the velocities/momenta at a given point in
space and time in order to compute the resulting, or emergent, velocity/momentum
field $\mathbf{v}_{i}=\frac{1}{m}\hbar\boldsymbol{\kappa}_{i},\:\emph{i}=\ensuremath{1}\mathrm{\:or\:}\ensuremath{2}$.
This, then, provides the correct intensity distributions and average
trajectories (lower plane). }\label{fig:interf-2}}
\par\end{center}%
\end{minipage}
\end{figure}

As in our model particles and fields are dynamically interlocked,
it would be highly misleading to picture the quantum potential in
a manner similar to the classical scenario of particle plus field,
where the latter can be switched on and off like an ordinary potential.
Contrariwise, in our case the particle velocities/momenta must be
considered as \emph{emergent}. One can illustrate this with the situation
in double-slit interference (Figure~\ref{fig:interf-2}). Considering
an incoming beam of, say, electrons with wave number $\mathbf{k}$
impinging on a wall with two slits, two beams with wave numbers $\mathbf{k}_{A}$
and $\mathbf{k}_{B}$, respectively, are created, which one may denote
as ``pre-determined'' quantities, resulting also in pre-determined
velocities $\mathbf{v}_{\alpha}=\frac{1}{m}\hbar\mathbf{k}_{\alpha}$,
$\alpha=\ensuremath{A}\:\mathrm{or}\:B$. 

However, if one considers that the electrons are not moving in empty
space, but in an undulatory environment created by the ubiquitous
zero-point field ``filling'' the whole experimental setup. One has
to combine all the velocities/momenta at a given point in space and
time in order to compute the resulting, or emergent, velocity/momentum
field $\mathbf{v}_{i}=\frac{1}{m}\hbar\boldsymbol{\kappa}_{i}$, $i=1\,\mathrm{or}\,2$
(Figure~\ref{fig:interf-2}), where $i$ is a bookkeeping index not
necessarily related to the particle coming from a particular slit~\cite{Fussy.2014multislit}.
The relevant contributions other than the particle's forward momentum
$m\mathbf{v}$ originate from the osmotic momentum $m\mathbf{u}$.
The latter is well known from Nelson's stochastic theory~\cite{Nelson.1966derivation},
but its identical form has been derived by one of us from an assumed
sub-quantum nonequilibrium thermodynamics~\cite{Groessing.2008vacuum,Groessing.2009origin}
as it was described above. As shall be shown in the next section,
our model also provides an understanding and deeper-level explanation
of the microphysical, causal processes involved, i.e.\ of the guiding
law~\cite{Groessing.2015implications} of the de~Broglie\textendash Bohm
theory. 

\section{Derivation of the De Broglie\textendash Bohm Guiding Equation for
\emph{$N$} Particles}

Consider at first one particle in an $n$-slit system. In quantum
mechanics, as well as in our emergent quantum mechanics approach,
one can write down a formula for the total intensity distribution
$P$ which is very similar to the classical formula. For the general
case of $n$ slits, it holds with phase differences $\varphi_{ii'}=\varphi_{i}-\varphi_{i'}$
between the slits $i$, $i'$ that
\begin{equation}
P=\sum_{i=1}^{n}\left(P_{i}+\sum_{i'=i+1}^{n}2R_{i}R_{i'}\cos\varphi_{ii'}\right),\label{eq:Sup2.1}
\end{equation}
where the phase differences are defined over the whole domain of the
experimental setup. As in our model the ``particle'' is actually
a bouncer in a fluctuating wave-like environment, i.e.~analogously
to the bouncers of the Couder experiments, one does have some (e.g.\ Gaussian)
distribution, with its centre following the Ehrenfest trajectory in
the free case, but one also has a diffusion to the right and to the
left of the mean path which is just due to that stochastic bouncing.
Thus the total velocity field of our bouncer in its fluctuating environment
is given by the sum of the forward velocity $\VEC v$ and the respective
osmotic velocities $\VEC u_{\mathrm{L}}$ and $\VEC u_{\mathrm{R}}$
to the left and the right. As for any direction $\alpha$ the osmotic
velocity $\VEC u_{\alpha}=\frac{\hbar}{2m}\frac{\nabla P}{P}$ does
not necessarily fall off with the distance, one has long effective
tails of the distributions which contribute to the nonlocal nature
of the interference phenomena~\cite{Groessing.2013dice}. In sum,
one has three distinct velocity (or current) channels per slit in
an $n$-slit system. 

We have previously shown~\cite{Fussy.2014multislit,Groessing.2014relational}
how one can derive the Bohmian guidance formula from our two-momenta
approach. Introducing classical wave amplitudes $R(\VEC w_{i})$ and
generalized velocity field vectors $\VEC w_{i}$, which represent
either a forward velocity $\VEC v$ or an osmotic velocity $\VEC u$
in the direction transversal to $\VEC v$, we calculate the phase-dependent
amplitude contributions of the total system's wave field projected
on one channel's amplitude $R(\VEC w_{i})$ at the point $(\VEC x,t)$
in the following way. We define a \emph{relational intensity} $P(\VEC w_{i})$
as the local wave intensity $P(\VEC w_{i})$ in each channel (i.e.~$\VEC w_{i}$),
recalling that there are 3 velocity channels per slit: $\VEC u_{\mathrm{L}}$,
$\VEC u_{\mathrm{R}}$, and $\VEC v$. The sum of all relational intensities,
then, is the total intensity, i.e.\ the total probability density.
In an $n$-slit system, we thus obtain for the relational intensities
and the corresponding currents, respectively, i.e.\ for each channel
component $\mathit{i}$,
\begin{align}
P(\VEC w_{i}) & =R(\VEC w_{i})\VEC{\hat{w}}_{i}\cdot{\displaystyle \sum_{i'=1}^{3n}}\VEC{\hat{w}}_{i'}R(\VEC w_{i'})\label{eq:Proj-1}\\
\VEC J\mathrm{(}\VEC w_{i}\mathrm{)} & =\VEC w_{i}P(\VEC w_{i}),\qquad i=1,\ldots,3n
\end{align}
with unit vectors $\VEC{\hat{w}}_{i}$ and
\begin{equation}
\cos\varphi_{ii'}:=\VEC{\hat{w}}_{i}\cdot\VEC{\hat{w}}_{i'}\,.
\end{equation}
Consequently, the total intensity and current of our field read as
\begin{align}
P_{\mathrm{tot}}= & {\displaystyle \sum_{i=1}^{3n}}P(\VEC w_{i})=\left({\displaystyle \sum_{i=1}^{3n}}\VEC{\hat{w}}_{i}R(\VEC w_{i})\right)^{2}\label{eq:Ptot6-1}\\
\VEC J_{\mathrm{tot}}= & \sum_{i=1}^{3n}\VEC J(\VEC w_{i})={\displaystyle \sum_{i=1}^{3n}}\VEC w_{i}P(\VEC w_{i}),\label{eq:Jtot6-1}
\end{align}
 leading to the \textit{emergent total velocity}
\begin{equation}
\VEC v_{\mathrm{tot}}=\frac{\VEC J_{\mathrm{tot}}}{P_{\mathrm{tot}}}=\frac{{\displaystyle \sum_{i=1}^{3n}}\VEC w_{i}P(\VEC w_{i})}{{\displaystyle \sum_{i=1}^{3n}}P(\VEC w_{i})}\,,\label{eq:vtot_fin-1}
\end{equation}
which represents the \textit{probability flux lines.}

In~\cite{Groessing.2012doubleslit,Fussy.2014multislit} we have shown
with the example of $n=2,$ i.e.\ a double-slit system, that Eq.~(\ref{eq:vtot_fin-1})
can equivalently be written in the form
\begin{equation}
\VEC v_{\mathrm{tot}}=\frac{R_{1}^{2}\VEC v_{\mathrm{1}}+R_{2}^{2}\VEC v_{\mathrm{2}}+R_{1}R_{2}\left(\VEC v_{\mathrm{1}}+\VEC v_{2}\right)\cos\varphi+R_{1}R_{2}\left(\VEC u_{1}-\VEC u_{2}\right)\sin\varphi}{R_{1}^{2}+R_{2}^{2}+2R_{1}R_{2}\cos\varphi}\,.\label{eq:vtot-1}
\end{equation}

The trajectories or streamlines, respectively, are obtained according
to $\VEC{\dot{x}}=\VEC v_{\mathrm{tot}}$ in the usual way by integration.
As we have first shown in~\cite{Groessing.2012doubleslit}, by re-inserting
the expressions for forward and osmotic velocities, respectively,
i.e.
\begin{equation}
\VEC v_{i}=\frac{\nabla S_{i}}{m}\,,\qquad\VEC u_{i}=-\frac{\hbar}{m}\frac{\nabla R_{i}}{R_{i}}\,,\label{eq:velocities}
\end{equation}
one immediately identifies Eq.~(\ref{eq:vtot-1}) with the Bohmian
guidance formula. Naturally, employing the Madelung transformation
for each slit $\alpha$ ($\alpha=1$ or $2$), 
\begin{equation}
\psi_{\alpha}=R_{\alpha}\e^{\mathrm{i}S_{\alpha}/\hbar},\label{eq:3.14-1}
\end{equation}
and thus $P_{\alpha}=R_{\alpha}^{2}=|\psi_{\alpha}|^{2}=\psi_{\alpha}^{*}\psi_{\alpha}$,
with $\varphi=(S_{1}-S_{2})/\hbar$, and recalling the usual trigonometric
identities such as $\cos\varphi=\frac{1}{2}\left(\e^{\mathrm{i}\varphi}+\e^{-\mathrm{i}\varphi}\right)$,
one can rewrite the total average current immediately in the usual
quantum mechanical form as 
\begin{equation}
\begin{array}{rl}
{\displaystyle \mathbf{J}_{{\rm tot}}} & =P_{{\rm tot}}\mathbf{v}_{{\rm tot}}\\ 
 & ={\displaystyle (\psi_{1}+\psi_{2})^{*}(\psi_{1}+\psi_{2})\frac{1}{2}\left[\frac{1}{m}\left(-\mathrm{i}\hbar\frac{\nabla(\psi_{1}+\psi_{2})}{(\psi_{1}+\psi_{2})}\right)+\frac{1}{m}\left(\mathrm{i}\hbar\frac{\nabla(\psi_{1}+\psi_{2})^{*}}{(\psi_{1}+\psi_{2})^{*}}\right)\right]}\\ 
 & ={\displaystyle -\frac{\mathrm{i}\hbar}{2m}\left[\Psi^{*}\nabla\Psi-\Psi\nabla\Psi^{*}\right]={\displaystyle \frac{1}{m}{\rm Re}\left\{ \Psi^{*}(-\mathrm{i}\hbar\nabla)\Psi\right\} ,}}
\end{array}\label{eq:3.18-1}
\end{equation}
where $P_{{\rm tot}}=|\psi_{1}+\psi_{2}|^{2}=:|\Psi|^{2}$.

Eq.~(\ref{eq:vtot_fin-1}) has been derived for one particle in an
$n$-slit system. However, for the spinless particles obeying the
Schrödinger equation\footnote{As we do not yet have a relativistic model involving spin our results
for the many-particle case cannot account for the difference in particle
statistics, i.e.\ for fermions or bosons. This will be a task for
future work.} it is straightforward to extend this derivation to the many-particle
case. Due to the purely additive terms in the expressions for the
total current and total probability density, respectively, also for
\emph{$N$} particles, Eqs.~(\ref{eq:Ptot6-1}) and (\ref{eq:Jtot6-1})
become
\begin{align}
P_{\mathrm{tot,}N} & =\sum_{j=1}^{N}\left[{\displaystyle \sum_{i=1}^{3n}}P(\VEC w_{i})\right]_{j}=\sum_{j=1}^{N}\left[\left({\displaystyle \sum_{i=1}^{3n}}\VEC{\hat{w}}_{i}R(\VEC w_{i})\right)^{2}\right]_{j}\,,\\
\VEC J_{\mathrm{tot,}N} & =\sum_{j=1}^{N}\left[\sum_{i=1}^{3n}\VEC J(\VEC w_{i})\right]_{j}=\sum_{j=1}^{N}\left[{\displaystyle \sum_{i=1}^{3n}}\VEC w_{i}P(\VEC w_{i})\right]_{j}\,,
\end{align}
and, analogously, Eq.~(\ref{eq:vtot_fin-1}),
\begin{align}
\VEC v_{\mathrm{tot,}N} & =\frac{\VEC J_{\mathrm{tot}}}{P_{\mathrm{tot}}}=\frac{{\displaystyle \sum_{j=1}^{N}}\left[{\displaystyle \sum_{i=1}^{3n}}\VEC w_{i}P(\VEC w_{i})\right]_{j}}{{\displaystyle \sum_{j=1}^{N}}\left[{\displaystyle \sum_{i=1}^{3n}}P(\VEC w_{i})\right]_{j}}\,,
\end{align}
where $\VEC w_{i}$ is dependent on the velocities~(\ref{eq:velocities})
with different $S_{i}$ and $R_{i}$ for every $j$. In quantum mechanical
terms the only difference now is that the currents' nabla operators
have to be applied at all of the locations of the respective \emph{$N$}
particles, thus providing 
\begin{equation}
{\displaystyle \mathbf{J}_{{\rm tot}}}\left(N\right)={\displaystyle \sum_{j=1}^{N}}\frac{1}{m_{j}}{\rm Re}\left\{ \Psi^{*}\left(t\right)(-\mathrm{i}\hbar\nabla_{j})\Psi\left(t\right)\right\} ,
\end{equation}
where $\Psi\left(t\right)$ now is the total $N$-particle wave function,
whereas the flux lines are given by
\begin{equation}
\VEC v_{j}\left(t\right)=\frac{\hbar}{m_{j}}\mathrm{Im}\frac{\nabla_{j}\Psi\left(t\right)}{\Psi\left(t\right)}\qquad\forall j=1,...,N.
\end{equation}

In sum, with our introduction of a relational intensity $P(\VEC w_{i})$
for channels $\VEC w_{i}$, which include sub-quantum velocity fields,
we obtain the guidance formula also for $N$-particle systems in real
3-dimensional space.\textsl{\emph{ The central ingredient for this
to be possible is to consider the emergence of the velocity field
from the interplay of the totality of all of the system's velocity
channels.}}

\begin{figure}[h]
\centering{}%
\begin{minipage}[t]{0.9\columnwidth}%
\begin{center}
\includegraphics[width=1.0\columnwidth,height=1.0\columnwidth]{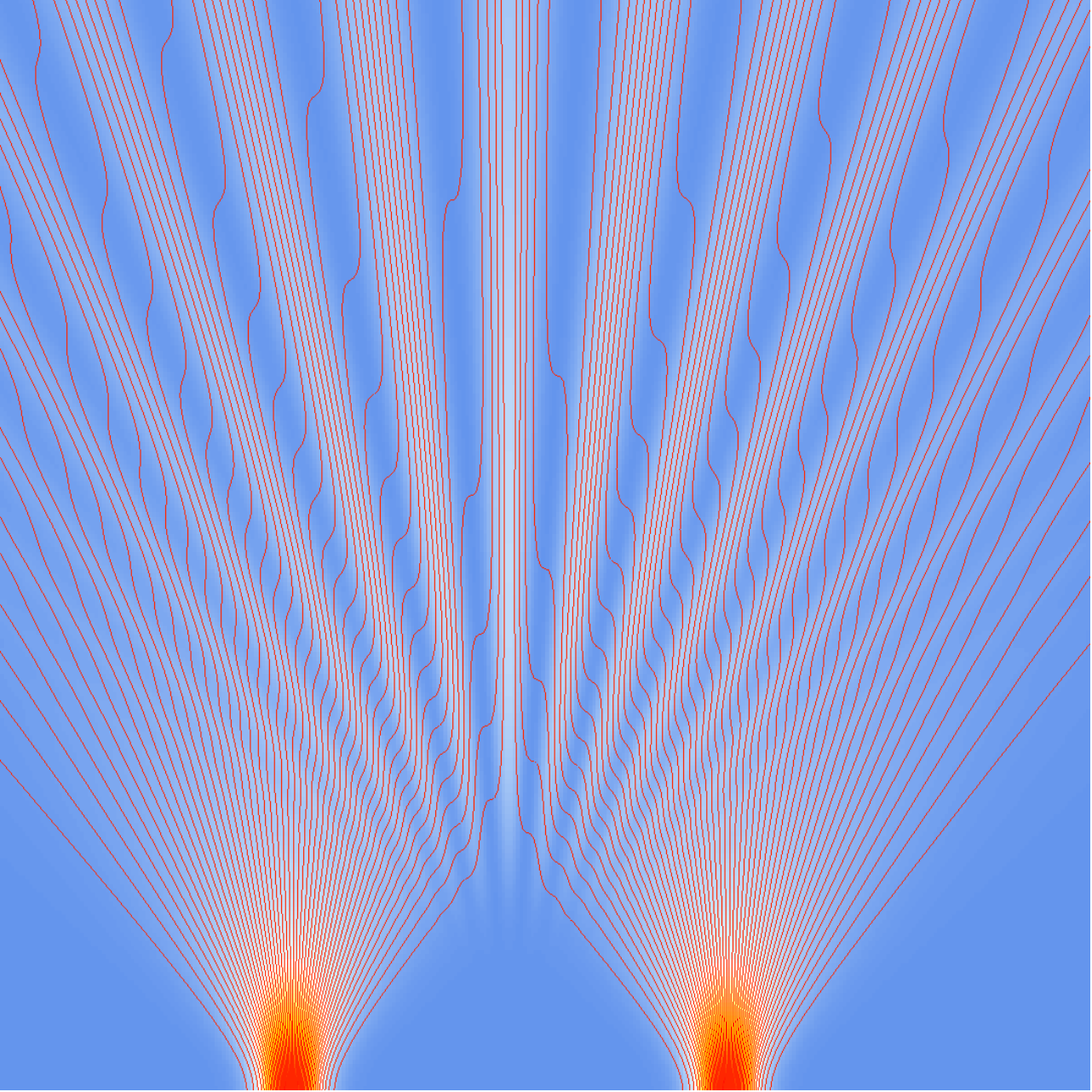}
\caption{Classical computer simulation of the interference pattern: intensity
distribution with increasing intensity from white through yellow and
orange, with trajectories (red) for two Gaussian slits, and with \textbf{large
dispersion} (evolution from bottom to top; $v_{x,1}=v_{x,2}=0$).
\label{fig:3}}
\par\end{center}%
\end{minipage}
\end{figure}
\begin{figure}[h]
\centering{}%
\begin{minipage}[t]{0.9\columnwidth}%
\begin{center}
\includegraphics[width=1.0\columnwidth,height=1.0\columnwidth]{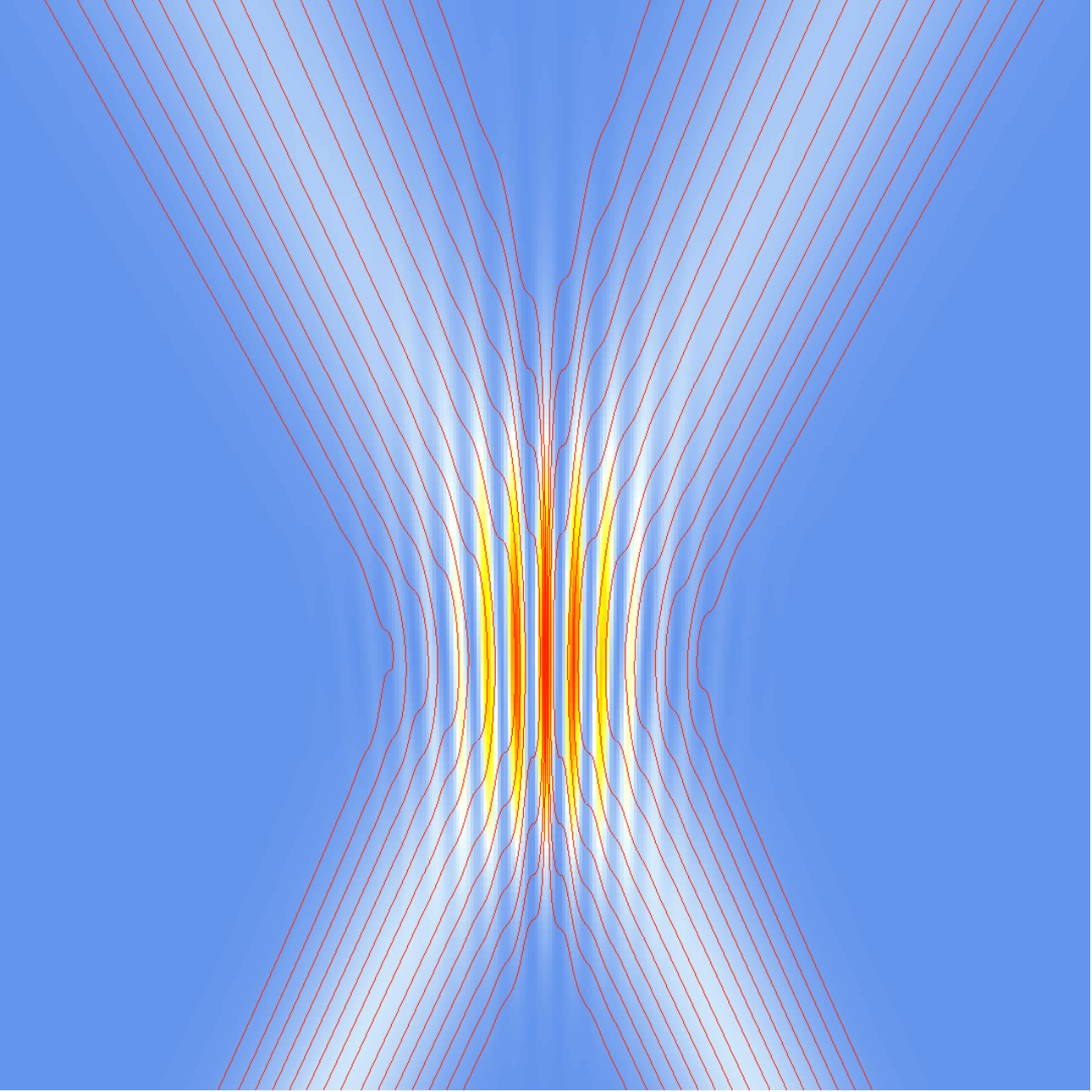}
\caption{Classical computer simulation of the interference pattern: intensity
distribution with increasing intensity from white through yellow and
orange, with trajectories (red) for two Gaussian slits, and with \textbf{small
dispersion} (evolution from bottom to top; $v_{x,1}=-v_{x,2}$). \label{fig:2}}
\par\end{center}%
\end{minipage}
\end{figure}

In Figures~\ref{fig:3} and \ref{fig:2}, trajectories (flux lines)
for two Gaussian slits are shown (from ref.~\cite{Groessing.2012doubleslit}).
These trajectories are in full accordance with those obtained from
the Bohmian approach, as can be seen by comparison with references~\cite{Holland.1993},
\cite{Bohm.1993undivided}, and \cite{Sanz.2009context}, for example.

\section{Vacuum Landscaping: Cause of nonlocal influences without signalling\label{sec:vacuum}}

In the foregoing sections, we pointed out how nonlocality appears
in our model. Particularly in discussing Eqs.~(\ref{eq:2.9})\textendash (\ref{eq:2.9c}),
it was shown that the form of the osmotic momentum 
\begin{equation}
m\mathbf{u}=-\frac{\hbar}{2}\frac{\nabla P}{P}
\end{equation}
 may be responsible for relevant influences. Moreover, if one assumes
a particle at some position $\VEC x$ in space, and with a probability
distribution $P$, the latter is a distribution around $\VEC x$ with
long tails across the whole experimental setup which may be very thin
but still non-zero. Then, even at locations $\VEC y$ very remote
from $\VEC x$, and although the probability distribution $P$ pertaining
to the far-away particle might be minuscule, it still may become effective
immediately through the zero-point field. 

The physical reason for bringing in nonlocality is the assumed resonant
coupling of the particle(s) with fluctuations of the zero-point vacuum
filling the whole experimental setup. Take, for example, a typical
``Gaussian slit''. We effectively describe $P$ by a Gaussian with
long non-zero tails throughout the whole apparatus. As we have seen,
in order to calculate on-screen distributions (i.e.\ total intensities)
of particles which went one-at-a-time through an $n$-slit device,
one only needs a two-momentum description and a calculation which
uses the totality of all relational intensities involving the relative
phases determined across the whole apparatus.

In general, we propose a resonant interaction of the bouncing ``particle''
with a \emph{relevant environment}.\footnote{In a similar vein, Bohm~\cite{Bohm.1980wholeness} speaks of a ``relatively
independent subtotality'' of the universe, to account for the possible
neglect of the ``rest of the universe'' in practical calculations.} For idealized, non-interacting particles, this relevant environment
would be the whole universe, and thus the idealized prototype of the
``cosmological solution'' referred to in the introduction.

For any particle in any experimental setup, however, the relevant
environment is defined by the boundary conditions of the apparatus.
Whereas the idealized one-particle scenario would constitute an indefinite
order of vibrations w.r.t. the particle oscillations potentially locking
in, the very building up of an experiment may represent a dynamical
transition from this indefinite order to the establishment of a definite
order. The latter is characterized by the emergence of standing waves
between the boundaries of the apparatus (like, e.g., source and detector),
to which the particle oscillations lock in. Moreover, if an experimenter
decides to change the boundary conditions (e.g., by altering the probability
landscape between source and detector), such a ``switching'' would
establish yet another definite order. The introduction or change of
boundary conditions, which immediately affects the probability landscape,
and the forward and the osmotic fields, we term ``vacuum landscaping''.

In other words, the change of boundary conditions of an experimental
arrangement constitutes the immediate transition from one cosmological
solution in the relevant environment (i.e.\ within the old boundary
conditions) to another (i.e.\ the new ones). The ``surfing'' bouncer/particle
just locally jumps from the old to the new standing wave solutions,
respectively. This is a process that happens locally for the particle,
practically instantaneously (i.e.\ within a time span $\propto\nicefrac{1}{\omega}),$
and nonlocally for the standing waves, due to the very definition
of the cosmological solutions. The vacuum landscape is thus nonlocally
changed without the propagation of ``signals'' in a communication
theoretical sense.\footnote{It is \emph{exclusively} the latter that must be prohibited in order
to avoid causal loops leading to paradoxes. See Walleczek and Grössing~\cite{Walleczek.2014non-signalling,Walleczek.2016nonlocal}
for an extensive clarification of this issue.}

We have, for example, discussed in some detail what happens in a double-slit
experiment if one starts with one slit only, and when the particle
might pass it, one opens the second slit~\cite{Groessing.2013dice,Groessing.2013joaopessoa}.
In accordance with Tollaksen \textit{et~al.}~\cite{Tollaksen.2010quantum}
we found that the opening of the second slit (i.e.\ a change in boundary
conditions) results in an uncontrollable shift in momentum on the
particle passing the first slit. Due to its uncontrollability (or,
the ``complete uncertainty'' in~\cite{Tollaksen.2010quantum}),
this momentum shift cannot be used for signalling. Still, it is necessary
to \textit{a posteriori} understand the final distributions on the
screen which would be incorrect without acknowledging said momentum
kick.

Similarly, Aspect-type experiments of two-particle interferometry
can be understood as alterations of vacuum landscapes. Consider, for
example, the case in two-particle interferometry, where Alice and
Bob each are equipped with an interfering device and receive one of
the counter-propagating particles from their common source. If Alice
during the time-of-flight of the particles changes her device by making
with suitable mirrors one of the interferometer arms longer than the
other, this constitutes an immediate switching from one vacuum landscape
to another, with the standing waves of the zero-point field now reflecting
the new experimental arrangement. In other words, the $P$-field has
been changed nonlocally throughout the experimental setup, and therefore
also all relational intensities
\begin{equation}
P(\VEC w_{i})=R(\VEC w_{i})\VEC{\hat{w}}_{i}\cdot{\displaystyle \sum_{i'}}\VEC{\hat{w}}_{i'}R(\VEC w_{i'})
\end{equation}
involved. The latter represent the relative phase shifts $\delta\varphi_{i,i'}=\delta\arccos\VEC{\hat{w}}_{i}\cdot\VEC{\hat{w}}_{i'}$
occurring due to the switching, and this change is becoming manifest
also in the total probability density
\begin{equation}
P_{\mathrm{tot}}={\displaystyle \sum_{i}}P(\VEC w_{i})=\left({\displaystyle \sum_{i}}\VEC{\hat{w}}_{i}R(\VEC w_{i})\right)^{2},
\end{equation}
with $i$ running through all channels of both Alice and Bob. The
quantum mechanical nonlocal correlations thus appear without any propagation
(e.g., from Alice to Bob), superluminal or other. As implied by Gisin's
group~\cite{Bancal.2012quantum}, this violates a ``principle of
continuity'' of propagating influences from \emph{A} to \emph{B},
but its non-signalling character is still in accordance with relativity
and the nonlocal correlations of quantum mechanics. Practically instantaneous
vacuum landscaping by Alice and/or Bob thus ensures the full agreement
with the quantum mechanical predictions without the need to invoke
(superluminal or other) signalling. Our model is, therefore, an example
of nonlocal influencing without signalling, which was recently shown
to provide a viable option for realistic modelling of nonlocal correlations.~\cite{Walleczek.2014non-signalling,Walleczek.2016nonlocal}

\section{Conclusions and outlook\label{sec:conclusion}}

With our two-momentum approach to an emergent quantum mechanics we
have shown that one can in principle base the foundations of quantum
mechanics on a deeper level that does not need wavefunctions. Still,
one can derive from this new starting point, which is largely rooted
in classical nonequilibrium thermodynamics, the usual nonrelativistic
quantum mechanical formalism involving wavefunctions, like the Schrödinger
equation or the de~Broglie\textendash Bohm guiding law. With regard
to the latter, the big advantage of our approach is given by the fact
that we avoid the troublesome influence from configuration space on
particles in real space, which Bohm himself has called ``indigestible''.
Instead, in our model the guiding equation is completely understandable
in real coordinate space, and actually a rather typical consequence
of the fact that the total current is the sum of all particular currents,
and the total intensity, or probability density, respectively, is
the sum of all relational intensities. As we are working with Schrödinger
(i.e.\ spinless) particles, accounting for differences in particle
statistics is still an open problem.

As shown, we can replicate quantum mechanical features exactly by
subjecting classical particle trajectories to diffusive processes
caused by the presence of the zero point field, with the important
property that the probability densities involved extend, however feebly,
over the whole setup of an experiment. The model employs a two-momenta
approach to the particle propagation, i.e., forward and osmotic momenta.
The form of the latter has been derived without any recurrence to
other approaches such as Nelson's.

The one thing that \emph{is} to be digested from our model is the
fact that the relational intensities are nonlocally defined, over
the whole experimental arrangement (i.e.\ the ``relevant environment'').
This lies at the bottom of our deeper-level ansatz, and it is the
\emph{only} difference to an otherwise completely classical approach.
We believe that this price is not too high, for we obtain a logical,
realistic picture of quantum processes which is rather simple to arrive
at. Nevertheless, in order to accept it, one needs to radically reconsider
what an ``object'' is. We believe that it is very much in the spirit
of David Bohm's thinking to direct one's attention away from a particle-centred
view and consider an alternative option: that the universe is to be
taken as a totality, which only under very specific and delicate experimental
arrangements can be broken down to a laboratory-sized relevant environment,
even if that laboratory might stretch along interplanetary distances.
In our approach, the setting up of an experimental arrangement limits
and shapes the forward and osmotic contributions and is described
as vacuum landscaping. Accordingly, any change of the boundary conditions
can be the cause of nonlocal influences throughout the whole setup,
thus explaining, e.g., Aspect-type experiments. We argue that these
influences can in no way be used for signalling purposes in the communication
theoretic sense, and are therefore fully compatible with special relativity.

Accepting that the vacuum fluctuations throughout the universe, or
at least within such a laboratory, are a defining part of a quantum,
amounts to seeing any object like an ``elementary particle'' as
nonlocally extended and, eventually, as exerting nonlocal influences
on other particles. For anyone who can digest this, quantum mechanics
is no more mysterious than classical mechanics or any other branch
of physics. 

\begin{turnpage}
\end{turnpage}

\begin{ruledtabular}
\end{ruledtabular}

\begin{turnpage}
\end{turnpage}

\appendix
\begin{acknowledgments}
We thank Jan Walleczek for many enlightening discussions, and the
Fetzer Franklin Fund of the John E. Fetzer Memorial Trust for partial
support of the current work. Also, we wish to thank Lev Vaidman and
several other colleagues for stimulating discussions. We thank the
latter for the exchange of viewpoints sometimes closely related to
our approach, as they also become apparent in their respective works:
Herman Batelaan~\cite{Batelaan.2016momentum}, Ana María Cetto and
Luis de la Pe{\~n}a~\cite{Cetto.2014emerging-quantum}, Hans-Thomas
Elze~\cite{Elze.2018configuration}, Basil Hiley~\cite{Hiley.2016structure},
Tim Maudlin~\cite{Maudlin.2011quantum}, Travis Norsen~\cite{Norsen.2016bohmian},
Garnet Ord~\cite{Ord.2009quantum}, and Louis Vervoort~\cite{Vervoort.2016no-go}.
\end{acknowledgments}

\providecommand{\href}[2]{#2}\begingroup\raggedright\endgroup

\end{document}